\documentstyle[amstex,amssymb,preprint,aps]{revtex}

\begin{document}
\title{Scaled Equation of State for Supercooled Water Near the Liquid-Liquid
Critical Point }
\author{D. A. Fuentevilla and M. A. Anisimov}
\address{Department of Chemical \& Biomolecular Engineering and Institute\\
for Physical Science \& Technology, \\
University of Maryland, College Park, MD 20742}
\date{\today}
\maketitle

\begin{abstract}
We have developed a scaled parametric equation of state to describe and
predict thermodynamic properties of supercooled water. \ The equation of
state, built on the growing evidence that the critical point of supercooled
liquid-liquid water separation exists, is universal in terms of theoretical
scaling fields and is shown to belong to the Ising-model class of
universality. The theoretical scaling fields are postulated to be analytical
combinations of the physical fields, pressure and temperature. The equation
of state enables us to accurately locate the \textquotedblleft Widom
line\textquotedblright\ (the locus of stability minima) and determine that
the critical pressure is considerably lower than predicted by computer
simulations.
\end{abstract}

Upon supercooling, water exhibits anomalous behavior with sharply increasing
heat capacity, isothermal compressibility, and the magnitude of negative
thermal expansivity \cite{PD:03}. A thermodynamically consistent view on the
global phase behavior of supercooled water was formulated by Poole et al. 
\cite{PP:92}. According to this view, the observed anomalies are associated
with density and entropy fluctuations diverging at a critical point of
liquid-liquid coexistence that terminates the line of first-order
transitions between two liquid aqueous phases: low-density liquid and
high-density liquid. This "second-critical-point" scenario is supported by
extensive Monte Carlo and molecular dynamic simulations \cite{SS:93}, by a
modified van der Waals model that includes hydrogen-bond contributions \cite%
{PP:94}, and by the limited but impressive experimental evidence \cite%
{OM:98,M:05}. An alternative interpretation of the phase behavior of water,
the "singularity-free" scenario, attributes the increase in response
functions upon supercooling, through a thermodynamic consistency argument,
to the existence of a negatively-sloped locus of density maxima in the P-T
plane \cite{SS:96}. However, the second-critical-point scenario seems more
plausible in view of the experimental evidence of a first-order phase
transition between two amorphous-water glasses \cite{PD:03,OM:98,M:05}. The
global phase diagram of water is shown in Fig. 1. Two remarkable features
make the second critical point in water phenomenologically different from
the well-known gas-liquid critical point. The negative slope of the
liquid-liquid phase transition line in the P-T plane means that the higher
density liquid water is the phase with larger entropy. A very large value of
the slope at the critical point (about 30 times larger than for the
vapor-liquid transition at the critical point) indicates the significance of
the entropy change with respect to the density change, and correspondingly,
the importance of the entropy fluctuations. However, the location of the
liquid-liquid critical point, especially the value of the critical pressure
is uncertain. The simulation data yields a variety of the critical-pressure
values, from negative pressures to 3.4 kbar \cite{PD:03,SH:97}. \ 

In this Letter, we present a scaled parametric equation of state to describe
and predict thermodynamic properties of supercooled water. \ The equation of
state is universal and belongs to the Ising-model class of universality. \
The equation of state enables us to accurately locate the critical point and
the \textquotedblleft Widom line\textquotedblright\ \cite{LX:05} (the locus
of stability minima and order-parameter fluctuation maxima). \ In
particular, we conclude that the critical pressure is considerably lower
than obtained by computer simulations and we predict thermodynamic
properties in the regions inaccessible to experiments. \ 

It is commonly accepted that the critical behavior of all fluids, simple and
complex, belong to the universality class of the three-dimensional Ising
model \cite{CD:76}. \ Water is not an exception \cite{MA:04}. \ Near the
critical point the critical (fluctuation-induced) part, $\Psi _{{\rm cr}}$,
of an appropriate field-dependent thermodynamic potential $\Psi $ is a
universal function of two scaling fields, \textquotedblleft ordering", $%
h_{1} $, and \textquotedblleft thermal", $h_{2}$ \cite{CD:76}:

\begin{equation}
\Psi _{{\rm cr}}\simeq h_{2}^{2-\alpha }f\left( \frac{h_{1}}{h_{2}^{\beta
+\gamma }}\right) ,  \label{eq1}
\end{equation}%
where $\alpha =0.109$, $\beta =0.326$, $\gamma =1.239$, are universal
exponents (interrelated as $\alpha +2\beta +\gamma =2$) in the scaling power
laws (as a function of $h_{2}$ at $h_{1}=0$) for the \textquotedblleft weak"
susceptibility, the order parameter and the \textquotedblleft strong"
susceptibility, respectively. \ The first derivatives of the thermodynamic
potential with respect to the scaling fields define two scaling densities,
the \textquotedblleft order parameter" $\phi _{1}=-\partial \Psi _{{\rm cr}%
}/\partial h_{1}$ and the \textquotedblleft thermal density" $\phi
_{2}=-\partial \Psi _{{\rm cr}}/\partial h_{2}$. \ Respectively, the second
derivatives define three susceptibilities, \textquotedblleft strong" $\chi
_{1}=\left( \partial \phi _{1}/\partial h_{1}\right) _{h_{2}}$,
\textquotedblleft weak" $\chi _{2}=\left( \partial \phi _{2}/\partial
h_{2}\right) _{h_{1}}$, and \textquotedblleft cross" $\chi _{12}=\left(
\partial \phi _{1}/\partial h_{2}\right) _{h_{1}}=\left( \partial \phi
_{2}/\partial h_{1}\right) _{h_{2}}$. \ The universal scaling function $f$
contains two system-dependent amplitudes that originate from the initial
Hamiltonian. \ In the mean-field approximation $(\alpha =0$, $\beta =1/2$, $%
\gamma =1)$, the critical part of the thermodynamic potential is represented
by a Landau expansion, $\Psi _{{\rm cr}}=\frac{1}{2}a_{0}h_{2}\phi _{1}^{2}+%
\frac{1}{4}u_{0}\phi _{1}^{4}-h_{1}\phi _{1}$, where the constants $a_{{\rm 0%
}}$ and $u_{{\rm 0}}$ (the same coefficients as in the initial Hamiltonian)
play the role of the mean-field system-dependent amplitudes.

To apply the universal expression (\ref{eq1}) for describing the
liquid-liquid critical point in supercooled water, we assume the scaling
fields are analytical combinations of physical fields, the pressure $P$ and
the temperature $T$:%
\begin{eqnarray}
h_{1} &=&a_{1}\Delta \hat{P}+a_{2}\Delta \hat{T}+a_{3}\Delta \hat{P}^{2}
\label{eq2} \\
h_{2} &=&b_{1}\Delta \hat{T}+b_{2}\Delta \hat{P}  \label{eq3}
\end{eqnarray}%
with $\Delta \hat{P}=\left( P-P_{{\rm c}}\right) /\left( \rho _{{\rm c}}RT_{%
{\rm c}}\right) $ and $\Delta \hat{T}=\left( T-T_{{\rm c}}\right) /T_{{\rm c}%
}$ (where the subscript \textquotedblleft {\rm c}" here and below indicates
the critical parameters) and $a_{i}$ and $b_{i}$ are system-dependent
coefficients. \ Representation of scaling fields through linear mixing of
physical fields is commonly used to incorporate asymmetric fluid criticality
into the symmetric Ising model \cite{NW:92,MA:98}. \ To account for the
strong curvature of the liquid-liquid transition line, defined as $h_{1}=0$,
we added a non-linear pressure term, $a_{3}\Delta \hat{P}^{2}$, in $h_{1}$.

Such a representation of the scaling fields corresponds to the practically
convenient choice of the molar Gibbs energy (chemical potential) as the
field-dependent potential so that $\Psi =\Psi (P,T)$ and $\Psi _{{\rm cr}%
}=\Psi _{{\rm cr}}(h_{1},h_{2})$. \ In this formulation we neglect the
difference between $-\Delta \hat{V}\equiv -\left( V-V_{{\rm c}}\right) /V_{%
{\rm c}}=\left( \rho -\rho _{{\rm c}}\right) /\rho $ and $\left( \rho -\rho
_{{\rm c}}\right) /\rho _{{\rm c}}$, a reasonable approximation for weakly
compressible liquids. \ Any two coefficients in the scaling fields may be
absorbed in two system-dependent amplitudes of the scaling function $f$, so
that we adopt\ $a_{2}=1$ and $b_{2}=-1$. \ The negative sign of $b_{2}$
indicates that the liquid-liquid phase separation in supercooled water
occurs with increase of pressure (Fig. 1), in contrast to the vapor-liquid
phase separation. \ The value of $a_{1}$, $a_{3}$, and $b_{1}$ can be
determined from the shape of the liquid-liquid first-order transition curve.
\ The \textquotedblleft Widom line" in the one-phase region, $h_{1}=0$, is
an analytical continuation of the liquid-liquid transition curve from ${\rm C%
}^{{\rm \shortmid }}$ to lower pressures and higher temperatures.

The mixing of physical fields into the scaling fields, defined by Eqs. (\ref%
{eq2}) and (\ref{eq3}), means that the order parameter is a combination of
molar entropy and molar volume. \ In the linear approximation 
\begin{eqnarray}
\phi _{1} &\equiv &\frac{b_{1}\Delta \hat{V}+b_{2}\Delta \hat{S}}{%
a_{2}b_{2}+\left( a_{1}\right) _{{\rm eff}}b_{1}},  \label{eq4} \\
\phi _{2} &=&\frac{a_{2}\Delta \hat{V}+b_{2}\Delta \hat{S}}{%
a_{2}b_{2}-\left( a_{1}\right) _{{\rm eff}}b_{1}},  \label{eq5}
\end{eqnarray}%
where $\Delta \hat{S}\equiv (S-S_{{\rm c}})/R$, with $R$ being the gas
constant, and $\left( a_{1}\right) _{{\rm eff}}=\left( \partial
h_{1}/\partial \hat{P}\right) _{\hat{T}}=a_{1}+2a_{3}\Delta \hat{P}$.

As far as the physical fields are mixed into the scaling fields, the
physical properties, such as the isobaric heat capacity $C_{P}$, the
isothermal compressibility $\kappa _{T}$, and the thermal expansivity $%
\alpha _{P}$,\ will not exhibit universal power laws when measured along
isotherms or isobars; instead, their apparent behavior will be determined
by\ a thermodynamic path and by the values of the mixing coefficients in
Eqs. (2) and (3). \ As follows from Eqs. (\ref{eq2}) and (\ref{eq3}), the
critical (fluctuation induced) parts of the dimensionless isobaric heat
capacity, isothermal compressibility, and thermal expansivity are expressed
through the scaling susceptibilities:%
\begin{eqnarray}
\left( \hat{C}_{P}\right) _{{\rm cr}} &=&\hat{T}\left( \frac{\partial \hat{S}%
}{\partial \hat{T}}\right) _{\hat{P}}-\left( \hat{C}_{P}\right) _{{\rm b}} \\
&=&-\hat{T}\left( a_{2}^{2}\chi _{1}+2a_{2}b_{1}\chi _{12}+b_{1}^{2}\chi
_{2}\right) ,  \nonumber \\
\left( \hat{\kappa}_{T}\right) _{{\rm cr}} &=&-\frac{1}{\hat{V}}\left( \frac{%
\partial \hat{V}}{\partial \hat{P}}\right) _{\hat{T}}-\left( \hat{\kappa}%
_{T}\right) _{{\rm b}} \\
&=&\frac{1}{\hat{V}}\left( \left( a_{1}\right) _{{\rm eff}}^{2}\chi
_{1}+2\left( a_{1}\right) _{{\rm eff}}b_{2}\chi _{12}+b_{2}^{2}\chi
_{2}\right) ,  \nonumber \\
\left( \hat{\alpha}_{P}\right) _{{\rm cr}} &=&\frac{1}{\hat{V}}\left( \frac{%
\partial \hat{V}}{\partial \hat{T}}\right) _{\hat{P}}-\left( \hat{\alpha}%
_{P}\right) _{{\rm b}} \\
&=&-\frac{1}{\hat{V}}\left( \left( a_{1}\right) _{{\rm eff}}a_{2}\chi
_{1}+\left( \left( a_{1}\right) _{{\rm eff}}b_{1}+a_{2}b_{2}\right) \chi
_{12}+b_{1}b_{2}\chi _{2}\right) \text{.}  \nonumber
\end{eqnarray}%
where $\hat{T}=T/T_{{\rm c}}$, $\hat{P}=P/\rho _{{\rm c}}RT_{{\rm c}}$, and
the subscript "{\rm b}" indicates the property backgrounds.

We use the simplest form of a scaled parametric equation of state, the
so-called \textquotedblleft linear model", which represents the scaling
fields and scaling susceptibilities as functions of the \textquotedblleft
polar" variables $r$ and $\theta $ \cite{MA:98,MS:00}: 
\begin{eqnarray}
h_{1} &=&ar^{\beta +\gamma }\theta \left( 1-\theta ^{2}\right) ,\
h_{2}=r\left( 1-b^{2}\theta ^{2}\right) , \\
\chi _{1} &=&\frac{k}{a}r^{-\gamma }c_{1}(\theta )\text{, }\chi
_{12}=kr^{\beta -1}c_{12}(\theta )\text{, }\chi _{2}=akr^{-\alpha
}c_{2}(\theta )-B_{{\rm cr}}  \nonumber
\end{eqnarray}%
where the coefficient $b^{2}=\left( \gamma -2\beta \right) /\gamma \left(
1-\beta \right) \simeq 1.36$ is a universal constant, while $a$ and $k$ are
system-dependent amplitudes, and $B_{{\rm cr}}$ is the so-called
\textquotedblleft critical background" of order $ak$ \cite{MA:98}. \ The
analytical functions $c_{1}(\theta )$, $c_{2}(\theta )$, and $c_{12}(\theta
) $ are calculated in ref. \cite{MA:98}. \ A remarkable feature of the
\textquotedblleft linear model" is that the singularities in the
thermodynamic functions are only related to the variable $r$, while the
properties are analytical with respect to $\theta $.

This model offers a consistent scaling description of the available
experimental data in supercooled water. \ Using high-resolution experimental
heat-capacity data \cite{A:82} shown in Fig. 2, we optimized the location of
the critical point and the system dependent amplitudes. \ Based on the most
recent estimate of the liquid-liquid phase transition curve given by Mishima 
\cite{M:05}, we have obtained the coefficients $a_{1}=b_{1}=0.0078$ $\ $and $%
a_{3}=0.062$. \ These particular numbers correspond to $P_{{\rm c}}=27$ MPa,
the value optimized by our equation of state. \ We assume that the
\textquotedblleft Widom line" is described by the same coefficients. \
Furthermore, to reduce the number of adjustable parameters, we assume that
the ratio $k/a=1$, as obtained for the three dimensional Ising model with
short range interactions \cite{MA:02}. \ Hence, only two adjustable
parameters, namely, $P_{{\rm c}}$ and $a=k$, have been used to describe the
anomalous parts of the thermodynamic properties. \ The non-critical
background of the heat capacity was approximated as a linear function of
temperature. \ We obtained $a=k=0.47$ and $P_{{\rm c}}=27$ MPa with the
critical temperature corresponding to this pressure $T_{{\rm c}}=232$ K. \
The critical point, obtained from our equation of state, is located at a
much lower pressure than previously predicted from computer simulations (see
Fig. 1a). \ 

With the given amplitudes and location of the critical point, we predict the
behavior of the compressibility and expansivity, shown in Figs. 3 and 4, by
adjusting only their non-critical backgrounds. \ The molar volume as a
function of temperature was taken from ref. \cite{Z:69}. \ The predictions
appear to have excellent agreement with the experimental data \cite%
{K:79,H:86}. \ While it is difficult to establish the error bars for the
obtained $P_{{\rm c}}$ value, the parametric equation of state certainly
excludes the critical pressure above 50 MPa or below 10 MPa. \ We also
conclude that the mean-field scenario is unlikely. \ The mean-field scenario
cannot predict the anomalous behavior of isothermal compressibility within
our model. \ While the major contribution in the heat-capacity anomaly is
strong susceptibility, $\chi _{1}$, ($b_{1}$ is small) diverging both in
mean-field and in scaling theory, the major contribution in the isothermal
compressibility anomaly is the weak susceptibility $\chi _{2}$, ($a_{1}$ is
small) which shows no anomaly in mean-field approximation. \ The major
contribution in the critical part of the expansivity comes from the cross
susceptibility $\chi _{12}$ as both $a_{1}$ and $b_{1}$ are small. \ These
features make the second critical point in water essentially different from
the liquid-vapor critical point where $C_{P}$, $\kappa _{T}$, and $\alpha
_{P}$ all diverge strongly, as $\chi _{1}$, and from the liquid-liquid
critical points in binary fluids where $C_{P}$, $\kappa _{T}$, and $\alpha
_{P}$ all diverge weakly, as $\chi _{2}$.

There are obvious limitations of our equation of state. \ First, the model
used in this work is accurate only asymptotically close to the critical
point $(r<<1)$ while all measurements in supercooled water have been taken
far beyond the asymptotic region. \ The experimental range of $r$, the
parametric distance to the critical point, may be as large as 0.5. \
However, this is the first estimate of the critical parameters for the
second critical point in water based on experimental data, and not on
computer simulations of "water like" models. \ Including non-asymptotic
corrections to the parametric equation of state would change the adjustable
backgrounds while not significantly affecting the critical parameters. \ To
more accurately describe and predict the properties in a broader range of
pressures and densities in supercooled water, a \textquotedblleft global"
crossover equation of state \cite{MS:00}, based on a reliable mean-field
equation of state, such as a modified van der Waals model \cite{PP:94}, is
required. \ Moreover, we did not address an intriguing possibility of the
existence of multiple critical points in supercooled water, as predicted by
some simulated water models \cite{VH:05}. \ 

In this work, the order parameter is phenomenologically expressed through
molar volume and entropy, with entropy being the major contribution. \ A
clarification of the relation between this phenomenology and the microscopic
nature of the order parameter \cite{TT:99}\ would help in better
understanding the physics of phase transitions in supercooled water. \ 

We acknowledge valuable discussions with P. Debenedetti, O. Mishima, H.E.
Stanley, and B. Widom.

\newpage

{\bf Captions of figures}

\bigskip

Figure 1. \ (a) Phase diagram for water with vapor-liquid and liquid-liquid
critical points. \ Solid curves are vapor-liquid, liquid-liquid, and
liquid-solid phase transitions; dashed is the Widom line; dotted is the
compressibility maxima; crosses are literature estimates for the
liquid-liquid critical point. \ {\rm C} is the vapor-liquid critical point; 
{\rm C}$^{{\rm \shortmid }}$ is the liquid-liquid critical point predicted
by our equation of state. \ (b) The vicinity of the liquid-liquid critical
point. \cite{OM:98,SH:97}.

Figure 2. \ Heat capacity measurements (stars) \cite{A:82} compared to the
heat capacity predicted by our scaling parametric equation of state (solid
curve) for a critical point of $P_{{\rm c}}=27$ MPa, $T_{{\rm c}}=232$ K. \
Non-critical background plotted as a thin solid curve.

Figure 3. \ Isothermal compressibility experimental data \cite{K:79} at 10
MPa (o), 50 MPa ($\triangle $), 100 MPa ($\Diamond $), 150 MPa ($\bigstar $%
), and 190 MPa ($\square $), compared with our prediction at the same
pressures (solid curves).

Figure 4. \ Thermal expansivity experimental data \cite{H:86} ($%
\blacktriangle $), at ambient pressure, compared with our prediction (solid
curve). \ Non-critical background plotted as a thin solid curve.


\bigskip

\begin{references}
\bibitem{PD:03} P.F. Debenedetti, J. Phys.: Condens. Matter {\bf 15}, R1669
(2003).

\bibitem{PP:92} P.H. Poole, F. Sciortino, U.Essmann, and H.E. Stanley,
Nature {\bf 360}, 324 (1992).

\bibitem{SS:93} S. Sastry, F. Sciortino, and H.E. Stanley, J. Chem. Phys. 
{\bf 98}, 9863 (1993); P.H. Poole, F. Sciortino, U. Essmann, and H.E.
Stanley, Pys. Rev. E {\bf 48}, 3799 (1993); P.H. Poole, U. Essmann, F.
Sciortino, and H.E. Stanley, Phys. Rev. E {\bf 48}, 4605 (1993); H.E.
Stanley, C.A. Angell, U. Essmann, M. Hemmati, P.H. Poole, and F. Sciortino,
Physica A {\bf 205}, 122 (1994).

\bibitem{PP:94} P.H. Poole, F. Sciortino, T. Grande, H.E. Stanley, and C.A.
Angell, Phys. Rev. Lett. {\bf 73} 1632 (1994).

\bibitem{OM:98} O. Mishima and H.E. Stanely, Nature {\bf 392}, 164 (1998).

\bibitem{M:05} O. Mishima, J. Chem. Phys. {\bf 23} 154506 (2005); O. Mishima
(personal communication).

\bibitem{SS:96} S. Sastry, P.G. Debenedetti, F. Sciortino, and H.E. Stanley,
Phys. Rev. E {\bf 53}, 6144 (1996); P.G. Debenedetti, Nature {\bf 392}, 127
(1998); L.P.N. Rebelo, P.G. Debenedetti, and S. Sastry, J. Phys. Chem. {\bf %
109}, 626 (1998).

\bibitem{SH:97} S. Harrington, R. Zhang, P.H. Poole, F. Sciortino, and H.E.
Stanley, Phys. Rev. Lett. {\bf 78}, 2409 (1997); F. Sciortino, P.H. Poole,
U. Essmann, and H.E. Stanley, Phys. Rev. E {\bf 55}, 727 (1997); M. Yamada,
S. Mossa, H.E. Stanley, and F. Sciortino, Phys. Rev. Lett. {\bf 88}, 195701
(2002); A. Scala, F.W. Starr, E. La Nave, H.E. Stanley, and F. Sciortino,
Phys. Rev. E {\bf 62}, 8016 (2000); H. Tanaka, Nature {\bf 380}, 328 (1996);
H. Tanaka, J. Chem. Phys. {\bf 105}, 5099 (1996).

\bibitem{LX:05} L. Xu, P. Kumar, S.V. Buldyrev, S.-H. Chen, P.H. Poole, F.
Sciortino, and H.E. Stanley, {\it Proc. Natl. Acad. Sci.} {\bf 102}, 16558
(2005).

\bibitem{CD:76} M.E. Fisher in {\it Critical Phenomena}, edited by F.J.W.
Hahne, Vol. 186 (Springer, Berlin, 1982), p. 1.

\bibitem{MA:04} M.A. Anisimov, J.V. Sengers, and J.M.H. Levelt Sengers, in 
{\it Near-Critical Behavior of Aqueous Systems}, edited by D.A. Palmer, R.
Fernandez-Prini, A.H. Harvey, The Physical Properties of Aqueous Systems at
Elevated Temperatures and Pressures: Water, Steam and Hydrothermal Solutions
(Academic Press, 2004).

\bibitem{NS:92} N.B. Wilding and A.D. Bruce, J. Phys.: Condensed Matter {\bf %
4}, 3087 (1992); M.E. Fisher and G. Orkoulas, Phys. Rev. Lett. {\bf 85}, 696
(2000); M.A. Anisimov and J.T. Wang, Phys. Rev. Lett. (in press).

\bibitem{MA:98} M.A. Anisimov, V.A. Agayan, and P.J. Collings, Phys. Rev. E, 
{\bf 57}, 582 (1998).

\bibitem{MS:00} M.A. Anisimov and J.V. Sengers, in {\it Equations of State
for Fluids and Fluid Mixtures}, edited by J.V. Sengers, R.F. Kayser, C.J.
Peters, and J.J. White Jr. (Elsevier, Amsterdam, 2000), p. 381.

\bibitem{A:82} C.A. Angel, W.J. Sichina, and M. Oguni, J. Phys. Chem. {\bf 86%
}, 998 (1982).

\bibitem{MA:02} Y.C. Kim, M.A. Anisimov, J.V. Sengers, and E. Luijten, J.
Stat. Phys. {\bf 110}, 591 (2003).

\bibitem{Z:69} B.V. Zheleznyi, Russ. J. Phys. Chem. {\bf 43}, 1311 (1969).

\bibitem{K:79} H. Kanno and C.A. Angel, J. Chem. Phys. {\bf 70}, 4008 (1979).

\bibitem{H:86} D.E. Hare and C.M. Sorensen, J. Chem. Phys. {\bf 84}, 5085
(1986).

\bibitem{VH:05} V.B. Henriques, N. Guisoni, M.A. Barbosa, M. Thielo, M.C.
Barbosa, Mol. Phys. {\bf 103}, 3001 (2005); I. Brovchenko, A. Geiger, A.
Oleinikova, J. Chem. Phys. {\bf 123}, 044515/1 (2005).

\bibitem{TT:99} T.M. Truskett, P.G. Debenedetti, S. Sastry, S. Torquato, J.
Chem. Phys. {\bf 111} 2647 (1999).
\end{references}
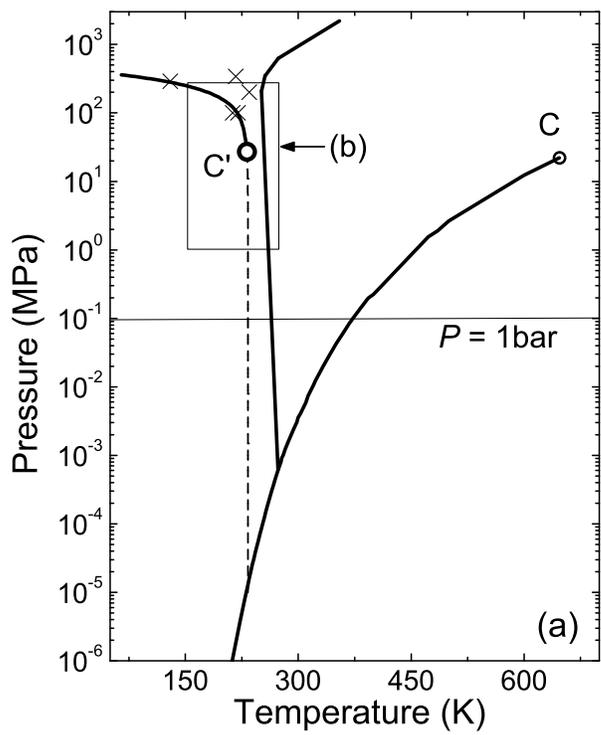
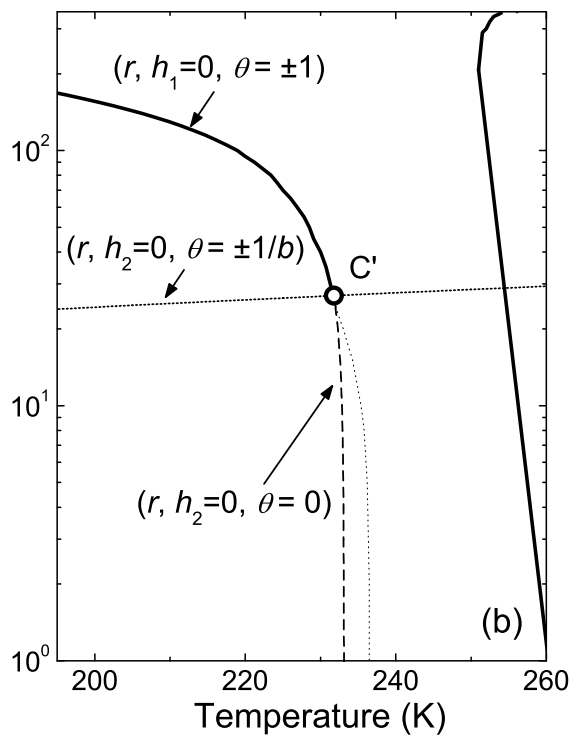
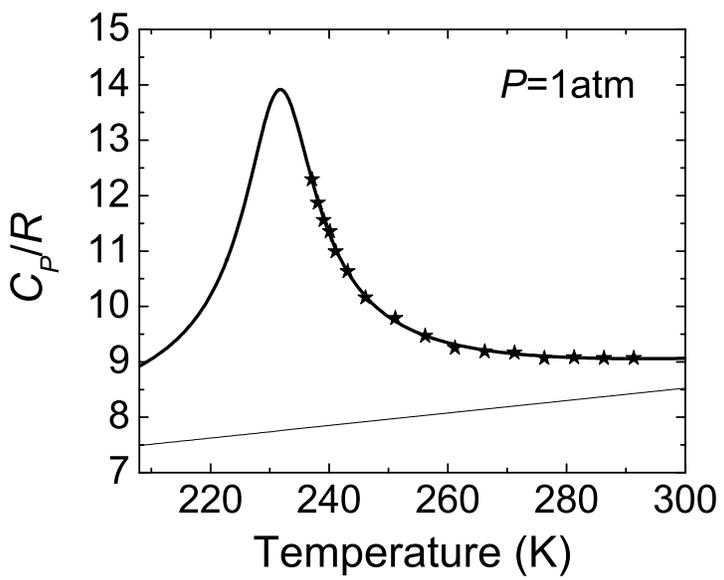
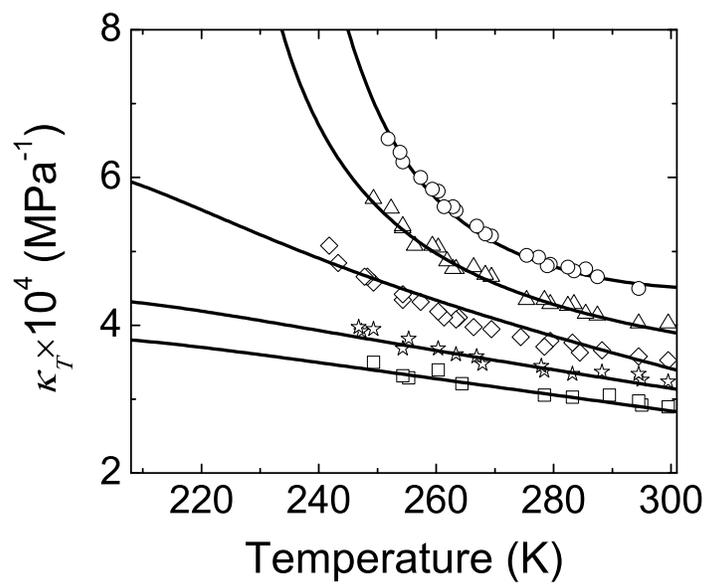
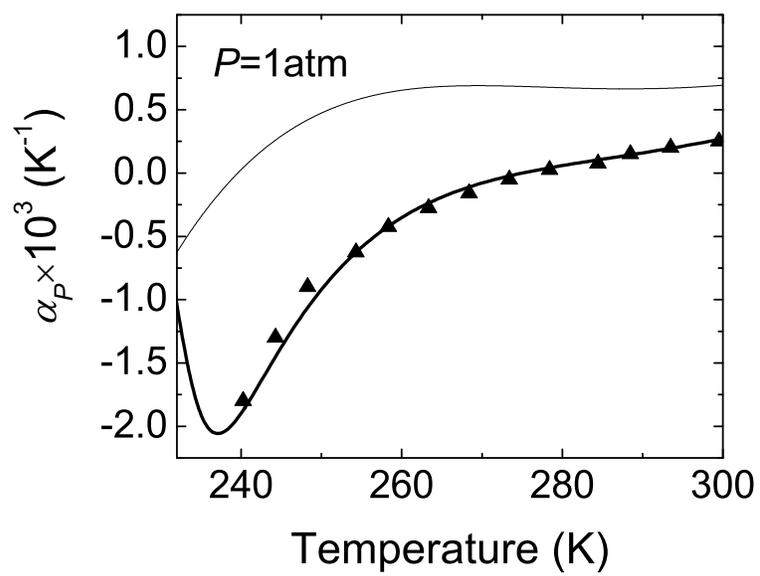
\end{document}